\documentclass[12pt]{article}

\newcommand{\dsl}
  {\kern.06em\hbox{\raise.15ex\hbox{$/$}\kern-.56em\hbox{$\partial$}}}

\newcommand{\be}{\begin{equation}}
\newcommand{\ee}{\end{equation}}
\newcommand{\ba}{\begin{eqnarray}}
\newcommand{\ea}{\end{eqnarray}}
\begin{document}
\title{Supersymmetric Dirac-Born-Infeld theory in 
noncommutative space}
\author{
N.~Grandi\thanks{Becario CONICET}\,
,  R.L.~Pakman$^*$\,  and 
F.A.~Schaposnik\thanks{Investigador CICBA}\\ {\normalsize\it
Departamento de F\'{\i}sica, Universidad Nacional de La Plata}\\
{\normalsize\it
  C.C.~67, (1900) La Plata, Argentina}%
}
\date{\hfill}
\maketitle
\begin{abstract}
We present a
supersymmetric version of Dirac-Born-Infeld (DBI) theory 
in noncommutative space-time, both
for  Abelian and non-Abelian gauge groups.
We show,
using the superfield formalism, 
 that the definition of a certain ordering with respect to the
$*$ product leads naturally to a DBI action both in the $U(1)$ as
well as in the $U(N)$ case. BPS equations are analysed in this context
and  properties of the resulting theory are discussed.
\end{abstract}

\maketitle
\newpage
\section{Introduction}

Noncommutative   gauge field theories have recently attracted 
much attention
after its connection with the
effective theory of strings was established \cite{cds}-\cite{dh}.
An explicit relation between  ordinary   and
noncommutative gauge theories was also obtained  and
 the equivalence of ordinary and noncommutative
Dirac-Born-Infeld (DBI) theory was proven  \cite{sw}. A supersymmetric
extension of the $U(1)$ DBI action when a $B$-field is 
turned on was analysed in \cite{fl} leading, in certain limit, to
the noncommutative version of the supersymmetric gauge theory.

When one defines  a DBI action in noncommutative
space, an ordering problem arises even in the $U(1)$ case. 
Indeed, since
multiplication of fields should be performed using the $*$ product, 
one necessarily faces a problem of
ordering when constructing a consistent  action; by this we mean  
defining an expansion in powers of 
$F_{\mu\nu}*F^{\mu\nu}$ and 
$F_{\mu\nu}*\tilde F^{\mu\nu}$
leading  to 
a square root DBI action.

If the interest in the DBI theory comes from D-brane dynamics, which
is relevant whenever derivative terms can be neglected, the ordering problem 
can be 
in principle ignored \cite{sw}. Nevertheless, the equivalence between
noncommutative and ordinary gauge theory through a change of
variables   imposes certain conditions on
derivative corrections to the DBI action, as it has
been analysed
in   refs. \cite{oka}-\cite{ot}. Other aspects of the noncommutative 
Dirac-Born-Infeld theory were discussed in refs. \cite{ga}-\cite{r}.

Another ordering problem concurrent with the one discussed above arises
for non-Abelian gauge symmetry. This second source of
 ordering problems is
related to the fact $F_{\mu\nu}$ takes values in the Lie algebra of $U(N)$.
Even in ordinary (commutative) space,  this last problem
has not been yet completely clarified. Indeed,  although the
symmetric trace proposed originaly in ref.\cite{Tse} seems to be the
natural operation to define a scalar DBI action 
and provides a good approximation to D-brane dynamics \cite{Tser},
disagreements with the full effective action
emerging from string theory have been signaled \cite{HT}. 

One advantage of using the symmetric trace to define 
the non-Abelian DBI action
in ordinary space is that it is the only one leading to an action which is
linearized by BPS conditions \cite{G1}-\cite{B}. This in turn  can be connected with
the possibility of supersymmetrizing the DBI action. 
Indeed, 
one can see that the natural superfield functionals
from which supersymmetric non-Abelian gauge theories are usually built,
combine  adequately giving the DBI form in such a way 
that the symmetric trace is singled out as the one to use in 
defining a scalar superfield DBI action
\cite{GSS}. Once the ($N=2$) supersymmetric DBI theory is constructed, BPS
relations can be derived from the SUSY algebra.

Following an approach similar to that in \cite{GSS}, we present here the
supersymmetric version of DBI theory in noncommutative space, both
for  Abelian and non-Abelian gauge groups. Using the superfield formalism, 
we shall see that definition of a certain ordering with respect to the
$*$ multiplication leads naturally to a DBI action both in the $U(1)$ as
well as in the $U(N)$ case. From the analysis of 
the supersymmetry transformations, BPS equations will be derived and the main
properties of the resulting theory will be discussed. 

The plan of the paper is
the following: the case of a $U(1)$ gauge theory in noncommutative 
space is discussed in section II. Imposing an adequate ordering
for superfields we show that a consistent supersymmetric DBI action can
be defined, this leading, in the bosonic sector, to a symmetric 
ordering
 of the DBI action. The extension to the non-Abelian $U(N)$ case
is presented in Section III. Finally, in Section IV we
discuss our results and also
analyse BPS equations  
in noncommutative space.

\section{The $U(1)$ case}

We  shall work in $3+1$ dimensional space-time with metric
$g^{\mu\nu} =(+,-,-,-)$  ($\mu,\nu =0,1,2,3$).
 The $*$ product between a pair of functions
$\varphi_1(x)$, $\varphi_2(x)$  is defined as
\begin{eqnarray}
(\varphi_1*\varphi_2)(x) &\equiv& \exp\left( \frac{i}{2} \theta_{ij}
\partial_{x_i}\partial_{y_j} \right) \left.
 \varphi_1(x)\varphi_2(y)\right\vert_{x=y}
\nonumber\\ &=& \varphi_1(x)\varphi_2(x) + \frac{i}{2} \theta_{ij}
\partial_i\varphi_1 \partial_j \varphi_1(x)
+ O(\theta^2)\; , \label{2}
\end{eqnarray}
Here, $\theta_{ij}$ ($i,j$ are spatial indices,  $i,j=1,2,3$) is a constant real-valued
antisymmetric tensor. The Moyal bracket is then defined as  
\begin{equation}
\{\varphi_1,\varphi_2\}(x) \equiv \varphi_1(x) *\varphi_2(x) - 
\varphi_2(x) * \varphi_1(x)\; ,
\label{4}
\end{equation}
so that, when applied to space-time  coordinates $x_\mu, x_\nu$, 
one has
\begin{equation}
\{x_\mu,x_\nu\} = i \theta_{\mu\nu} \label{1}
\end{equation}
which is why one refers to noncommutative spaces.
 We have
included in (\ref{1}) the time coordinate $x_0$,
 but we shall take  $\theta_{0i}=0 $ so that
definition (\ref{2}) will suffice.

A ``noncommutative gauge theory'' is defined just by using the
$*$ product each time the gauge fields are multiplied.
Then, even in the $U(1)$ Abelian case, the curvature $F_{\mu\nu}$
has a non-linear term (with the same origin as the usual
commutator in non-Abelian gauge theories in  ordinary space)

\begin{eqnarray}
F_{\mu\nu} &=& \partial_\mu A_\nu - \partial_\nu A_\mu - i
\left(A_\mu * A_\nu - A_\nu * A_\mu\right) \nonumber\\ &=&
\partial_\mu A_\nu - \partial_\nu A_\mu -
i\{A_\mu,A_\nu\}\; .
\label{5}
\end{eqnarray}
The field strength is gauge-covariant (even
in the Abelian case) under gauge transformations which should be
represented by ${\cal U}$ elements of the form
\begin{equation}
{\cal U}(x) = \exp_*(i\alpha) \equiv 1 + i \alpha - \frac{1}{2}
\alpha*\alpha + \ldots \label{6}
\end{equation}
The covariant derivative implementing infinitesimal gauge
transformations takes the form
\begin{equation}
{\cal D}_\mu[A] \alpha = \partial_\mu \alpha + i \left( \alpha
*A_\mu - A_\mu*\alpha\right) \label{7}
\end{equation}
so that an infinitesimal gauge transformation on $A_\mu$ reads as
usual
\begin{equation}
\delta A_\mu =  {\cal D}_\mu \alpha \label{in}
\end{equation}
Concerning finite gauge transformations, one has
\begin{equation}
A_\mu^{U} = {i} {\cal U}(x) * \partial_\mu {\cal U}^{-1}(x) + 
{\cal U}(x) *
A_\mu * {\cal U}^{-1}(x) \label{gt}
\end{equation}

Given a two-component spinor $\lambda$ one can see that the combination
\be
{\cal D}_\mu[A] \lambda =\partial_\mu \lambda+ i \left( \lambda
*A_\mu - A_\mu*\lambda\right) \label{rep}
\end{equation}
transforms covariantly under gauge transformations provided $\lambda$ 
transforms according to 
\begin{equation}
\lambda \to {\cal U}(x) *\lambda * {\cal U}^{-1}(x) \label{gt2}
\end{equation}

Superfields in noncommutative geometry have been discussed in refs.
\cite{fl},\cite{cz}-\cite{ter2}. We  take
the $*$ product between superfields $U_1, U_2, \ldots$ in the form
\begin{eqnarray}
(U_1*U_2)[x,\bar \theta,\theta] &\equiv& 
\exp\left( \frac{i}{2} \theta_{\mu\nu}
\partial_{x_\mu}\partial_{y_\nu} \right) \left.
 U_1[x,\bar \theta,\theta] U_2[y,\bar \theta,\theta]\right\vert_{x=y}
\label{gracias}
\end{eqnarray}
Note that the $*$ product involves only space-time coordinates without
affecting Grassmann coordinates $\theta$.

Consider a real vector superfield $V$ in the Wess-Zumino gauge
\be
V[x,\bar \theta,\theta] = - \theta^\alpha
 \sigma^\mu_{\,\alpha \dot \beta} \bar \theta^{\dot \beta}
  A_\mu + i \theta^\alpha \theta_\alpha \bar \theta_{\dot \beta}
\bar \lambda^{\dot \beta}  - i \bar \theta_{\dot \beta} 
\bar \theta^{\dot\beta}  \theta^\alpha
\lambda_\alpha + \frac{1}{2} \theta^\alpha \theta_\alpha 
\bar\theta_{\dot \beta} \bar \theta ^{\dot \beta}D
\label{V}
\ee
where $D$ is the auxiliary field. 

It is convenient to define  chiral variable $y^\mu$ and ${y^\dagger}^\mu$
in the form
\be
y^\mu = x^\mu + i \theta^\alpha \sigma^\mu_{\,\alpha \dot \beta} \bar 
\theta^{\dot \beta} \, ,   \;\;\;\;\;
{y^\dagger}^\mu = x^\mu - i \theta^\alpha \sigma^\mu_{\,\alpha \dot \beta} \bar 
\theta^{\dot \beta}
\label{t}
\ee
so that one can  define  derivatives $D_\alpha$ and $\bar D_\alpha$ as

\be
D_\alpha = ~ \frac \partial {\partial \theta ^\alpha }+2i
\left( \sigma ^\mu\bar
\theta \right) _\alpha \frac \partial {\partial y^\mu}, \ \ \ \ \
\bar D_{\dot\alpha} = -\frac{\partial}{\partial \bar\theta^{\dot\alpha}}
\label{4c}
\ee
when acting on functions of $(y,\theta,\bar\theta)$ and
\be
D_{\alpha} =  \frac{\partial}{\partial \theta^\alpha},
\ \ \ \
\bar D_{\dot \alpha } = - \frac
\partial {\partial \bar \theta ^{\dot \alpha }%
}-2i\left( \theta \sigma ^\mu\right) _{\dot \alpha }
\frac \partial {\partial y^{\dagger\mu}}
\label{5c}
\ee
on functions of $(y^\dagger,\theta,\bar\theta)$.
Generalized
gauge transformation acting on superfields will be written in the form
\be
\exp_*(2i\Lambda)=  1 + 2i\Lambda - 2 \Lambda*
\Lambda + \ldots
\label{ele}
\ee
where $\Lambda = \Lambda(y,\theta)$ is a chiral left-handed superfield and
$\Lambda^\dagger
(y^\dagger,\bar \theta)$
its right-handed Hermitian conjugate,
\be
\bar D_{\dot \alpha} \Lambda = D_\alpha \Lambda^\dagger = 0
\label{condi}
\ee
Explicitly,
\be
\Lambda(y,\theta) =
A  + \sqrt 2 \theta^\alpha \chi_\alpha + \theta^\alpha
\theta_{\alpha} G  
\label{ex}
\ee
where $A$ and  $G$ are complex scalar fields and $\chi$ is
a Weyl spinor.
Under such a transformation, the superfield
$V$ transforms as
\be
\exp_*(2V) \to \exp_*(-2i\Lambda^\dagger)* \exp_*(2V) * \exp_*(2i\Lambda)
\label{12b}
\ee
{}From $V$, the   chiral curvature 
superfield $W_\alpha$ can be constructed,
\begin{equation}
W^\alpha \left( y,\theta \right) =
-\frac{1}{8}
\bar D_{\dot\alpha} \bar D^{\dot\alpha} \exp_*(-2 V) * D^\alpha \exp_*(2 V)
\label{2ddddddd}
\end{equation}
In contrast with (\ref{12b}), under a gauge transformation $W_\alpha$
transforms covariantly,
\be
W^\alpha \to \exp_*(-2i\Lambda) * W^\alpha * \exp_*(2i\Lambda)
\label{12c}
\ee
Concerning the hermitian conjugate, $\bar W_\alpha$, it transforms as
\be
\bar W_{\dot \alpha} \to \exp_*(-2i\Lambda^\dagger)
*\bar W_{\dot\alpha} * \exp_*(2i\Lambda^\dagger)
\label{12d}
\ee
Written in components, $W_\alpha$ reads
\begin{equation}
W^\alpha \left( y,\theta \right) =-i\lambda ^\alpha
+\theta^\alpha D
-\frac{i}{2}\left( \sigma ^\mu\bar \sigma ^\nu \theta \right) ^\alpha
F_{\mu\nu}
+\theta \theta \left( \not\!\!{\cal D} \bar
\lambda \right) ^\alpha
\label{6c}
\end{equation}

Now, the $N=1$ supersymmetric Maxwell Lagrangian in noncommutative
space-time can be
written in terms of the chiral superfield $W^\alpha$ and its hermitian conjugate
$\bar W_{\dot \alpha}$ as 
\be
L_0 =\frac{1}{4e^2} \left(
\int d^2\theta W^\alpha * W_\alpha
+
\int d^2\bar \theta \bar W_{\dot \alpha} * \bar W^{\dot \alpha}
\right)
\ee
since the last components in $W*W$ and $\bar W* \bar W$ contain
the combination $D*D-\frac{1}{2}\left( F_{\mu \nu}* F^{\mu \nu}\pm
\right.$ $\left. i F_{\mu \nu}*\tilde F^{\mu \nu}\right)$. 
In order to construct a DBI-like Lagrangian in
noncommutative space, one should include
invariants which cannot be expressed in terms of the
basic quadratic invariants $F_{\mu\nu}*F^{\mu\nu}$
and $F_{\mu\nu}*\tilde F^{\mu\nu}$. Indeed, if one is to
search for a DBI Lagrangian expressible as a determinant, there will
appear Lorentz invariants which cannot be written in terms of the two
quadratic ones referred above. This is in fact the
case of the quartic terms which will necessarily arise
when computing a DBI determinant in four dimensions: 
although in commutative space these terms can be written as a
certain functional of $F_{\mu\nu}F^{\mu\nu}$ and 
$F_{\mu\nu}\tilde F^{\mu\nu}$,  the $*$ product prevents such
an accomodation. Moreover, odd powers which were absent
in the commutative space case could be now present.

Let us start at this point the search of a candidate for the supersymmetric
extension of the DBI model in noncommutative space.  We know
that higher powers of $W$ and
$\bar W$ should be included in such a way as to respect gauge and Lorentz
invariance. In the case of commutative space this was achieved by
combining $W^2 \bar W^2$ with adequate powers of $D^2W^2$
and $\bar D^2 \bar W^2$ \cite{DP}-\cite{GNSS}. In the present 
case, there are two sources of complications. First, 
in view of the transformation
laws (\ref{12b}) and (\ref{12c})-(\ref{12d}), $\exp_*(2V)$ factors should
be adequately included in quartic superfield terms in order to ensure gauge invariance.
Second, one has to include as many independent superfield
invariants as necessary
to reproduce those quartic Lorentz invariants which cannot be written
in terms of $F_{\mu\nu}*F^{\mu\nu}$ and $F_{\mu\nu}*\tilde F^{\mu\nu}$.

Consider then the possible gauge-invariant superfields
that can give rise to quatric terms. There are two natural candidates,
\ba
Q_1 &=& \!\int\! d^2\theta d^2 \bar \theta~
W^\alpha * W_\alpha *\exp_*({-2V}) * \bar W_{\dot \beta} * \bar
W^{\dot \beta}
* \exp_*({2V}) 
\nonumber\\
Q_2 &=& \int\! d^2\theta d^2 \bar \theta ~
W^\alpha  *
\exp_*({-2V}) *
\bar W^{\dot\beta} *
\exp_*({2V}) \nonumber\\
&&*
W_\alpha *
\exp_*({-2V}) *
\bar W_{\dot \beta}*
\exp_*({2V})
\ea

Now, the bosonic part of $Q_1$ is given by
\ba
& &\left. Q_1 \right \vert_{bos}
 =
D^{*4}
-\frac{1}{2}
\left\{
F^{\mu\nu}*F_{\mu\nu},~D^{*2}
\right\}_+
-\frac{i}{2}
\left\{
F^{\mu\nu}*\tilde F_{\mu\nu},~D^{*2}
\right\}
\nonumber\\
&&
+~
\frac 1 4
\left(
( F^{\mu\nu}*F_{\mu\nu})^{*2}
+
(F^{\mu\nu}*\tilde F_{\mu\nu})^{*2}
\right)
+
\frac i 4
\left\{F^{\mu\nu}*\tilde
F_{\mu\nu},~F^{\rho\sigma}*F_{\rho\sigma}\right\}
\label{-q1}
\ea
where $A^{*2} = A * A$. Then,
its purely $F_{\mu\nu}$ dependent part takes the form
\ba
Q_1|_{bos,\,F_{\mu\nu}}
\!\!\!&=& \!\!\!
\frac 1 4
\left(
       (F^{\mu\nu}*F_{\mu\nu})^{*2}
      +
       (F^{\mu\nu}*\tilde F_{\mu\nu})^{*2}
\right)
+
\frac i 4
\left\{
F^{\mu\nu}*\tilde
F_{\mu\nu},F^{\rho\sigma}*F_{\rho\sigma}
\right\} \nonumber\\
\label{1u}
\ea
Analogously, one gets for  $Q_2$,
\begin{eqnarray}
Q_2|_{bos,\,F_{\mu\nu}}
&= &\frac{1}{4} F^{\mu\nu}*F^{\rho\sigma}*F_{\mu\nu}*F_{\rho\sigma}
+
\frac{1}{4}\tilde F^{\mu\nu}*\tilde F^{\rho\sigma}*F_{\mu\nu}*F_{\rho\sigma}
\nonumber\\
& & +
\frac{i}{4}\left\{
F^{\mu\nu},F^{\rho\sigma} *
\tilde F_{\mu\nu}* F_{\rho\sigma}
\right\}
\label{2u}
\end{eqnarray}
One can now see that a very economic combination of $Q_1$ and $Q_2$ generates
the quartic terms to be expected in the expansion of a square
root DBI action, provided the latter is defined using a symmetric 
ordering of the $F$ factors. Indeed,
using that
\be
\int d^4x \{\varphi_1,\varphi_2\} =  0
\ee
one has
\ba
\frac{1}{4!}\int d^4x&& \!\!\!\!\!\!\!\!\left. \left(
2Q_1 + Q_2\right)\right\vert_{bos,\,F_{\mu\nu}}
 = \nonumber\\
 & &\int \!\! d^4x \, 
 {\cal S}^*
 \left. 
 \left(1 - 
_*\sqrt{1 + \frac{1}{2} F_{\mu\nu}*F^{\mu\nu}
- \frac{1}{16}\left(F_{\mu\nu}* \tilde F^{\mu\nu}\right)^2 }\,
\right)
\right\vert ^{4{\rm th~order}} 
\label{look}
\ea
where in the r.h.s. we have defined the symmetric ordering operation 
${\cal S}^*$
as
\be
 {\cal S}^* \left(
\varphi_1*\varphi_2*\ldots * \varphi_N
\right) = 
\frac{1}{N!} \sum_{\{\pi\}} \varphi_{\pi(1)}*\varphi_{\pi(2)}*\ldots * 
\varphi_{\pi(N)}
\label{ar}
\ee
In  eq.(\ref{look}) $_*\sqrt{}$ means that the square root power expansion
is defined using the $*$ product to build each power.

This symmetric ordering defined by eq.(\ref{ar})
ensures that products of real fields like the r.h.s. in
(\ref{look}) are real numbers, which is not guaranteed for a general
non ordered $*$ product.
 Also, ${\cal S}^*$ solves the ambiguities arising
in the definition of the DBI Lagrangian as a determinant and one
can prove that 
\ba
\int d^4x &&\!\!\!\!\!\!\!\!\!\!\!\!\!{\cal S}^*
\left(1 - _*\sqrt{1 + \frac{1}{2} (F_{\mu \nu}*F^{\mu \nu})
-\frac{1}{16} (F_{\mu \nu}*\tilde F^{\mu \nu})^2}\,
\right) \nonumber\\
&=& \int d^4x \,{\cal S}^*\left(1 -  
_*\sqrt{\det(g_{\mu \nu} + F_{\mu \nu})}
\right)
\label{chuqui}
\ea
with the r.h.s. univoquely defined through the ${\cal S}^*$ prescription. 
Then, if one arrives, through the supersymmetric construction 
to a bosonic action which takes the form of the l.h.s. in (\ref{chuqui})
one is sure that this action can be written also in the usual BI action
determinantal form.

The analysis above was made for the purely bosonic sector. It is then natural
to extend it by considering the complete superfield combination 
$\int d^4x(2Q_1 +Q_2)$
that again accomodates in the form of a symmetric ordering but which now
has to be defined so as to take into account the presence of
Grassmann anticommuting objects
\begin{eqnarray}
\frac{1}{3} \int d^4x 
(2 &&\!\!\!\! \!\!\!\! Q_1 +    Q_2)  = \nonumber\\
& & \int d^4x {\cal S}^*  \left(
W^\alpha*W_\alpha* \left[\exp_*(-2V)* \bar W_{\dot \beta} *
\exp_*(2V)\right]*
\right.\nonumber\\
&&
\left.
\left[\exp_*(-2V)*\bar W^{\dot \beta} *
\exp_*(2V)\right]
\right)
\label{sr2}
\end{eqnarray}
where  now the symmetric ordering  is defined as
\be
{\cal S}^* \left(
U^{1}*U^{2} 
 *\ldots * U^{n}
\right)  =
\frac{1}{n!}
 \sum_{\{\pi\}} {\rm sgn}\,\pi_O\, U^{{\pi(1)}}*
U^{{\pi(2)}}*\ldots * 
U^{{\pi(n)}}
\label{arq}
\end{equation}
Here $U$ represents either $W$ or $[\exp_*(-2V) \bar W \exp_*(2V)]$
and
 ${\rm sgn}\,\pi_O $ only takes care  of the sign of the permutation of  
Grassmann odd objects in each term.

Note that in eq.(\ref{sr2}) adyacent $\exp_*(2V)$ and $\exp_*(-2V)$ 
inside the ${\cal S}^*$ symbol cannot
be cancelled {\it before} the symmetric ordering is performed since
it  includes terms in which those factors are not adyacent. That
is, the expressions inside brackets in (\ref{sr2}) should be taken 
as individual elements for permutations.

Now, in order to get higher powers of $F_{\mu\nu}*F^{\mu\nu}$ and 
$F_{\mu\nu}*\tilde F^{\mu\nu}$
necessary to construct the DBI action, 
it has been
shown  \cite{DP}
that one has to include powers of some
superfields which we shall call
$X$ and $Y$ that in the present case should be defined as
\begin{eqnarray}
X \!\!\!&=& \!\!\!
\frac{1}{16}
\left(\left\{\bar D_{\dot \beta} \bar D^{\dot \beta}
\bar R_{\dot \alpha},\bar R^{\dot \alpha}
\right\}_+
-2
\bar D_{\dot \beta} \bar R_{\dot\alpha}
* 
\bar D^{\dot \beta} \bar R^{\dot\alpha} 
\right.
\nonumber\\
\!\!\!& & \!\!\!+ \left\{ \exp_*(-2V) * D^\beta D_\beta R^\alpha *
 \exp_*(2V), W_\alpha
\right\}_+
\nonumber\\
\!\!\!& & \!\!\!
-2\left.
\left[
{\rm exp}_*(-2V) *
D^{\beta} R^\alpha *
{\rm exp}(2V)
\right]*
\left[ 
{\rm exp}(-2V) *
D_{\beta} R_\alpha *
{\rm exp}(2V)
\phantom{D_{\dot \beta}}
\!\!\!\!\!\!\!\!
\right]
\right)
\nonumber\\
\nonumber\\
Y\!\!\!&=& \!\!\!
-\frac{i}{32}
\left(\left\{\bar D_{\dot \beta} \bar D^{\dot \beta}
\bar R_{\dot \alpha},\bar R^{\dot \alpha}
\right\}_+ 
-2
\bar D_{\dot \beta} \bar R_{\dot\alpha}
* 
\bar D^{\dot \beta} \bar R^{\dot\alpha} 
\right.
\nonumber\\
\!\!\!& & \!\!\!- \left\{ \exp_*(-2V) * D^\beta D_\beta R^\alpha *
 \exp_*(2V), W_\alpha
\right\}_+
\nonumber\\
\!\!\!& & \!\!\!
+2\left.
\left[
{\rm exp}_*(-2V) *
D^{\beta} R^\alpha *
{\rm exp}(2V)
\right]*
\left[ 
{\rm exp}(-2V) *
D_{\beta} R_\alpha *
{\rm exp}(2V)
\phantom{D_{\dot \beta}}
\!\!\!\!\!\!\!\!
\right]
\right)
\nonumber\\
& & \label{y}
\end{eqnarray}
where we have defined  
\begin{eqnarray}
\bar R_{\dot \alpha} = 
{\rm exp}_*(-2V)*\bar W_{\dot \alpha}*{\rm exp}_*(2V) \nonumber\\
R^{\alpha} = {\rm exp}_*(2V)*W^{\alpha}*{\rm exp}_*(-2V) 
\end{eqnarray}

In particular, one has in the bosonic sector
\be
X |_{\theta = \bar \theta = 0} =
-\frac{1}{2}D*D + \frac{1}{4} F^{\mu \nu} * F_{\mu \nu}
\label{uno}
\ee
\be
Y|_{\theta = \bar \theta = 0} = \frac{1}{8}
F^{\mu \nu}*\tilde F _{\mu \nu} 
\label{dos}
\ee
Hence,  lowest components of $X$ and $Y$ include the invariants 
$F_{\mu\nu}*F^{\mu \nu}$ and
$F_{\mu\nu}*\tilde F^{\mu \nu}$.

We are thus lead to consider the following supersymmetric action
as a candidate  to reproduce the DBI dynamics in its bosonic sector,
\newpage
\begin{eqnarray}
{\rm S}_{S}  &=& \int d^4x L_{0} + 
\left( \sum_{n,m} C_{nm} \int d^4x d^4\theta {\cal S}^*
\left(W^\alpha *
W_\alpha *  
 \bar R_{\dot \beta}\!
*\!
\bar R^{\dot \beta}*X^{*n}*Y^{*m}
\right) \right.\nonumber\\
& &  +\left.\phantom{\sum_{\pi}^7}\!\!\!\!
{\rm h.c.} \right)
\label{L}
\end{eqnarray}
Operation ${\cal S}^*$ was defined in eq.(\ref{arq}) for  
superfields $U$. This amounts, for $X$ and $Y$ factors, to 
order the (Grassmann even)  brackets entering in their construction
without change of sign.   

Expression (\ref{L}) gives a general action, the supersymmetric extension
of a bosonic gauge invariant action depending on the field strength 
$F_{\mu\nu}$
through the  quantities $F_{\mu\nu}*F^{\mu\nu}$ and 
$F_{\mu\nu}*\tilde F^{\mu\nu}$, in certain combinations constrained by
supersymmetry. Coefficients $C_{mn}$ remain arbitrary and can be chosen
so as to obtain the DBI action in its usual  form. Indeed, if
one puts
\begin{eqnarray}
 C_{0\,0}  & = & \frac{1}{8}
\nonumber\\
  C_{n-2m \,2m} & = & {(-1)^m} \sum_{j=0}^{m} \left (\matrix{
n+2-j\cr
j\cr
}\right ) q_{n+1-j}
\nonumber\\
  C_{n \,2m+1} & = & 0 ~,
\label{coe}
\end{eqnarray}
\begin{eqnarray}
q_0 &=&  -\frac{1}{2}
\nonumber\\
q_n & = & \frac{(-1)^{n+1}}{4n} \frac{(2n-1)!}{(n+1)!(n-1)!}  ~ ~ ~ {\rm for} ~
{}~ n\geq 1
\label{q}
\end{eqnarray}
then, the purely bosonic part of  action
(\ref{L}) becomes, after using relation (\ref{chuqui}),
\be
\left. S_S\right\vert_{bos}
= \int d^4x {\cal S}^* \left( 1 - _*\sqrt{-\det(g_{\mu \nu} + F_{\mu \nu})}
\right)
\label{LL}
\ee
As in the ordinary space case, other choices of coefficients $C_{mn}$ are
possible leading to consistent  supersymmetric gauge theories with 
non-polynomial gauge field dynamics. Let us note that odd powers of the
field strength $F_{\mu\nu}$ were excluded in our treatment since it is
not possible to construct a superfield functional of $W$, $\bar W$, 
$DW$ and $\bar D \bar W$ containing $F^3$ terms in its highest
 $\theta$ component.

\section{The $U(N)$ case}
The construction of the non-Abelian supersymmetric DBI action closely follows
the steps described in the previous section for the $U(1)$ case. One starts
from a real vector superfield $V$ as defined in (\ref{12}) but now with
the gauge field $A_\mu$, Majorana field $\lambda$ and auxiliary field $D$
taking values in the Lie algebra of $U(N)$,
\be
A_\mu = A_\mu^a t^a \;\;\; \;\;\;
\lambda = \lambda ^a t^a \;\;\; \;\;\; D = D^a t^a
\label{10}
\ee
with $t^a$ the (hermitian) $U(N)$ generators,
\be
[t^a,t^b] = i f^{abc} t^c
\label{11}
\ee
\be
{\rm tr} \, t^a t^b =  \delta^{ab}
\label{12}
\ee
Generalized gauge transformations are implemented as in eq.(\ref{ele})
with the chiral superfield $\Lambda = \Lambda^a t^a$. Analogously, superfields
$W^\alpha$ and $\bar W_{\dot \alpha}$ now take values in the Lie algebra
of $U(N)$.

In ordinary commutative space, the scalar superfield Lagrangian for Yang-Mills
theory is easily constructed by taking the trace tr  of  $W^2 +\bar W^2$. Now,
when one needs higher powers of combinations of $W$ and $\bar W$ to generate
DBI-like Lagrangians, an ordering problem associated with generators $t^a$
arises. Supersymmetry signals out a natural trace operation which solves
the ordering problem \cite{GSS}. This trace is precisely the symmetric
trace operation Str
introduced in \cite{Tse} in order to connect a DBI theory with the 
effective string action for non-Abelian vector fields. In the 
present noncommutative non-Abelian case, one faces a second ordering problem
related to the $*$ product. As we shall see both problems are entangled
and have to be solved both at once.

Consider the following operation on $n$ Lie algebra valued bosonic objects 
$F_1, F_2,  \ldots, F_n $ (for example $F_1 = F_{\mu\nu}^at^a $ ),
\be
{\rm tr}\, {\cal S}^* (F_1,F_2, \ldots ,F_n) = {\rm tr} 
\left(\frac{1}{n!}\sum_{\pi}  
F_{\pi_1}*F_{\pi_2}* \ldots *F_{\pi_n}\right)
\label{ge}
\ee
Note that the operation  ${\rm tr}\, {\cal S}^*$ coincides with the 
symmetric trace  Str in the $\theta_{\mu\nu} \to 0$ commutative limit. 
It is easy to see that (\ref{ge}) changes covariantly under gauge transformations ${\cal U}$  when $F_i \to
 {\cal U}^{-1}*F_i*{\cal U}$,   
\ba
 {\rm tr}\, {\cal S}^* (&&\!\!\!\!\!\!\!\!\!
 \!\!\!\![{\cal U}^{-1}* F_1*{\cal U}]*[{\cal U}^{-1}*F_2*
 {\cal U}] * \ldots *[{\cal U}^{-1}*F_n*{\cal U}])
\nonumber\\ 
&=& 
{\rm tr} 
\left(\frac{1}{n!}\sum_{\pi}  
{\cal U}^{-1}*F_{\pi_1}*{\cal U}*
{\cal U}^{-1}*F_{\pi_2}*{\cal U}* \ldots {\cal U}^{-1}*F_{\pi_n}*
{\cal U}\right)
\nonumber\\
&=& {\rm tr} 
\left(\frac{1}{n!}\sum_{\pi}  
{\cal U}^{-1}*F_{\pi_1}*F_{\pi_2}* \ldots F_{\pi_n}*{\cal U}\right)\nonumber\\
& =&{\rm tr} \, {\cal U}^{-1}* {\rm tr}\, {\cal S}^* (F_1,F_2, \ldots ,F_n) * 
{\cal U}
\label{ge2}
\ea
Formula (\ref{ge}) can then be used to construct powers of $F_{\mu\nu}$ leading
to a gauge covariant Lagrangian. 

Let us note that
instead of taking the normal trace  tr on Lie algebra
generators after the ${\cal S}^*$ operation, one could try to
use the
symmetric trace Str, 
symmetrizing further the product of generators ${t^a}'s$
without affecting the already symmetrized
coefficients $F_i^a$.  Then, one can see  that a different
expression (that also reduces to the symmetric trace prescription 
in the commutative limit) is obtained. However, this recipe for an
ordered product should be 
ruled out  since 
it lacks the gauge covariance property.

The generalization of formula (\ref{ge}) to a   product   of
Lie algebra valued superfields
is the same as in the $U(1)$ case (eq.(\ref{arq})). One defines,
\be
{\rm tr}  \,{\cal S}^* \left(
U_1*U_2*\ldots * U_n
\right) = 
\frac{1}{n!} {\rm tr} 
 \sum_{\{\pi\}} {\rm sgn}\,\pi_O \, U_{\pi(1)}*U_{\pi(2)}*\ldots * 
U_{\pi(n)}
\label{arq2}
\ee
Since the $U$'s are Lie algebra valued superfields,
the trace has been included in order to define a scalar
object.   

We are now ready to propose the non-Abelian generalization of the
DBI supersymmetric action in the form
\begin{eqnarray}
{\rm S}_{S}\!\!\!  &=&\!\!\!\!{\rm tr}\!\! \int \!\!d^4x d^2\theta   \,
W^\alpha * W_\alpha
  \nonumber\\
& & 
\!\!\!\! + \sum_{n,m} C_{nm} {\rm tr}\!\! \int \!\! d^4x d^4\theta\, {\cal S}^*
\left(W^\alpha *
W_\alpha *   \bar R_{\dot \beta}\!
*\! 
\bar R^{\dot \beta}*X^{*n}*Y^{*m}\right)
 + 
{\rm h.c.} 
\label{L2}
\end{eqnarray}
where $X$ and $Y$ are constructed from eqs.(\ref{y}) with
$W$, $\bar W$  and $V$ now in the Lie algebra of $U(N)$. The same 
choice
of coefficients $C_{mn}$ as in eqs.(\ref{coe})-(\ref{q}) leads now to the
non-Abelian noncommutative DBI action in the bosonic sector
\ba
\left. S_S\right\vert_{bos} &=&
\int d^4x  \, {\rm tr}\, {\cal S}^* 
\left(1 - _*\sqrt{1 + \frac{1}{2} (F_{\mu \nu}*F^{\mu \nu})
-\frac{1}{16} (F_{\mu \nu}*\tilde F^{\mu \nu})^2}\,
\right) \nonumber\\
&=& \int d^4x \, {\rm tr}\,
{\cal S}^* \left( 1 - _*\sqrt{-\det(g_{\mu \nu} + F_{\mu \nu})}
\right)
\label{LLL}
\ea
\section{Discussion}
In this work we have constructed the supersymmetric noncommutative
Born-Infeld action both for Abelian and non-Abelian gauge theories.
In our construction, we have seen that the natural superfield 
functionals from which SUSY extensions of gauge theories can be
constructed, combine in the adequate, square root DBI form in
such a way that a symmetric order ${\cal S}^*$ with respect to the $*$ product
is singled out. 

In the Abelian case, the supersymmetric action 
defined by eq.(\ref{L}) leads to a bosonic sector with  dynamics
governed by action (\ref{LL}) which then provides a consistent definition
of noncommutative $U(1)$ DBI action to all orders in $\theta_{\mu\nu}$.
The same construction can be performed for the $U(N)$ gauge theory.
Remarkably, in this case the problem of defining a scalar action from
Lie algebra valued objects reduces to simply taking the ordinary
trace of  ${\cal S}^*$-symmetrized products of fields. As we have shown, 
this prescription leads to  DBI action (eqs.(\ref{L2})-(\ref{LLL}))
which is gauge invariant, in contrast with  
what happens if one were to use the symmetric trace  after the ${\cal S}^*$
operation. 
Nevertheless, in the $\theta_{\mu\nu} \to 0$ limit, because of
the symmetrizing effect of ${\cal S}^*$, our
result coincides with the one obtained for the commutative non-Abelian 
DBI action defined using the symmetric trace.

An important aspect in the study of supersymmetric DBI actions 
concerns the BPS structure in the bosonic sector.  In  this respect,
and for commutative DBI theories, conditions under which the resulting
BPS  relations coincide with those arising 
in  Yang-Mills gauge theories have been studied and
their relation with $N=2$ supersymmetry clarified
 \cite{G1}-\cite{B}, \cite{CNS}.
In the noncommutative case, solutions to
different self-dual equations have been already found \cite{ns}-\cite{KLY}
but the analysis of the connection between those equations
and supersymmetry is lacking. It is then natural
to investigate this issue in the supersymmetric
framework presented here extending our treatment from 
 $N=1$  to the $N=2$ supersymmetry since it is in this last  case that
the BPS analysis can be performed. 

In order to achieve the $N=2$ extension,  one has to add to the
vector multiplet $( A_\mu,\lambda,D)$ already introduced, a chiral scalar 
multiplet $\Phi = (\phi,\psi,F)$ where $\phi=(\phi_1 + i \phi_2)$ is a complex scalar field, $\psi$ a
second Majorana fermion and $F$ an auxiliary complex field. A complete $N=2$ vector 
multiplet  can be accomodated in terms of all these fields in the form
$(A_\mu,\chi,\phi,D,F)$ where now $\chi$ is a Dirac fermion constructed from
$\lambda$ and $\psi$,  $\chi=(\lambda,\bar \psi)$. 

The $N=2$ DBI Lagrangian is constructed
following the same steps described for the $N=1$ case. One uses 
as building block
the $N=2$ chiral superfield $\Psi$ which plays a role analogous to the $N=1$
curvature superfield $W$,
\be
\Psi = \Phi(\tilde y,\theta_1) + i\sqrt 2  {\theta_2}^\alpha 
W_\alpha(\tilde y,\theta_1) + \theta_2^\alpha {\theta_2}_\alpha
G(\tilde y,\theta_1)
\ee
Here $\tilde y^\mu  = y^\mu + i \theta_2 \sigma^\mu \bar \theta_2$ and $G$
is a chiral scalar which can be expressed in terms of $\Phi$ and $V$ 
(see for example \cite{Ly}). 

It is easy to see that 
\begin{eqnarray}
\left.\Psi^2\right\vert_{\psi=\lambda=\phi=0} &=& -4 \theta_1^2 \theta_2^2
\left(F_{\mu\nu}*F^{\mu\nu}
+  i F_{\mu\nu}*\tilde F^{\mu\nu} - \frac{1}{2} D*D -\frac{1}{4}
  \{F^\dagger, F\}_+ \right)
\nonumber\\
&=& \theta_1^2 \theta_2^2 \, Z[F_{\mu\nu},D,F]
\end{eqnarray} 
and hence the bosonic part of the $N =2$  Lagrangian will depend on
$F_{\mu\nu}$  $D$ and $F$ through the combination defined by $Z$.  Then,
the equations of motion for $D$ and $F$ are solved by $D=F=0$.  

In order
to look for instanton configurations one has to pass to Euclidean space
and  equate
to zero the $N=2$ gaugino supersymmetry transformation law. Since
we have assumed $\theta_{0i} = 0$, the Wick rotation can be  
performed without any complication related to the definition of the
$*$ product. Now, 
in the gaugino transformation law  
all fermionic fields and the scalar field
$\phi$ should be put to zero while  
the auxiliary fields $D$ and $F$ should be eliminated using 
their equations of motion which, as we have seen, lead to $D=F=0$. 

The off-shell gaugino 
supersymmetry transformation takes the form
\be
\delta \chi =-i \left(\Gamma^{\mu\nu}F_{\mu\nu} - D\right)\xi + i\sqrt 2
\not \!\partial
\left(\gamma_5 \phi_1 + i  \phi_2\right)  \chi  - \sqrt 2 
F^\dagger \gamma_5 \bar \chi^t
\ee
where we have written
$(\xi_1,-\bar \xi_2) = \xi$ for the two   
parameters associated with $N=2$ supersymmetry
and $\Gamma^{\mu\nu} = (i/4)[\gamma^\mu,\gamma^\nu]$. 
Putting $\phi = D=F = 0$, the gaugino transformation law reduces to
\be
\delta \chi = -i \Gamma^{\mu\nu}F_{\mu\nu} \xi
\ee
so that, when imposing the BPS condition in the form $\delta \chi = 0$ one
obtains the self-duality equation which, now in Euclidean space, 
takes the form
\be
F_{\mu\nu} = \pm \tilde F_{\mu\nu}
\label{au}
\ee
Each one of the solutions breakes half of the supersymmetries. As expected,
the instanton equation formally coincides with that arising in ordinary
space-time. However, it should be remembered that $F_{\mu\nu}$ in 
eq.(\ref{au}) as defined in (\ref{5}) 
includes a Moyal bracket. This means that even in the $U(1)$ case
there could be instanton solutions which in fact have been found in \cite{ns}.
In order to get a deeper understanding of BPS equations and bound 
one should construct the $N=2$ supersymmetry algebra. We hope to discuss this
problem in a future work.

~

\noindent{\bf Acknowledgements}:
We would like to thank Guillermo Silva
for helpful discussions and comments.
This work is  supported in part by CICBA, CONICET (PIP
4330/96), and ANPCYT (PICT 97/2285) Argentina.


\end{document}